\documentclass[9pt]{article}
\usepackage{amsmath,amssymb,latexsym,enumerate}
\usepackage{epsfig,graphicx,float}
\usepackage{xcolor}

\newcommand{\be}{\begin{equation}}
\newcommand{\ee}{\end{equation}}
\newcommand{\sL}{\mathsf L}
\newcommand{\sT}{\mathsf T}
\newcommand{\sM}{\mathsf M}
\newcommand{\sI}{\mathsf I}
\newcommand{\sTh}{\mathsf\Theta}

\newcommand{\tmag}{t_{\text{mag}}}
\newcommand{\theat}{t_{\text{heat}}}
\newcommand{\tdif}{t_{\text{diff}}}

\newcommand{\tableref}[1]{\hyperref[tab:#1]{\textsc{Table} \ref*{tab:#1}}}
\newcommand{\figref}[1]{\hyperref[fig:#1]{\textsc{Figure} \ref*{fig:#1}}}

\newcommand{\Deltx}{\Delta x}
\newcommand{\Deltt}{\Delta t}

\begin{document}

\title{Mathematical analysis of a flux-jump model in superconductivity}
\author{Jean-Guy Caputo, Nathan Rouxelin} 
\maketitle

{\normalsize \noindent Laboratoire de Math\'ematiques, INSA de Rouen Normandie,\\
Normandie Universit\'e \\
76801 Saint-Etienne du Rouvray, France\\
E-mail: jean-guy.caputo@insa-rouen.fr, nathan.rouxelin@insa-rouen.fr \\

\date{\ }
\tableofcontents

{\bf Abstract} \\
Type II superconductors can trap a transient magnetic field
and become {\it cryomagnets} that are very useful for
applications. During this process, flux jumps i.e. sudden
jumps of the total magnetization occur
and hinder the properties of these magnets.
To understand the electrodynamics of these systems and in
particular flux jumps, we analyzed mathematically a model
based on Maxwell's equations and temperature in a 1D configuration.
When a magnetic pulse is applied to a superconductor, three effects
occur, from fastest to slowest: Joule heating, magnetic relaxation and
temperature diffusion.
Adimensionalising the problem, we obtain a nonlinear
diffusion for the magnetic field coupled to a forced diffusion
equation for the temperature with only two parameters.
Two regimes occur, depending on temperature: for medium temperature
the heat capacity of a sample can be assumed constant while
for low temperature it
depends on temperature causing a nonlinear temperature evolution.
Flux jumps can be explained using the fixed points of the equations.
We found that they occur for
pulses of duration close to the magnetic relaxation time and
mostly at low temperature because of the nonlinear dependance.
Flux trapping is maximal for medium amplitude long duration pulses
and low to medium temperatures, so these conditions are optimal
to produce better cryomagnets.

\parskip 0.2 cm 


\tableofcontents

\section{Introduction}

High T$_c$ superconductivity has permitted the advent of strong and flexible
superconducting magnets that can be used for screening magnetic fields
in medical instruments and build ultra light motors for transportation
and power, see the road map by Durrell et al \cite{roadmap} for all
the possible applications. The main technique to
fabricate a magnet, called {\it Pulse Field Magnetization} \cite{roadmap}, 
is to submit a cooled sample to a strong magnetic pulse.
During the process, a current is created inside the superconductor
and can remain fairly constant for weeks as long as the sample
remains cooled. Another phenomena can occur called {\it flux jump}, where 
the total magnetization of the sample varies suddenly, breaks the 
superconductivity and changes the properties of the magnet, see 
the experiments \cite{romero07}, \cite{fracasso23}, 
\cite{ainslie16} done with MgB$_2$ and YBaCuO materials.

A microscopic description of the electrodynamics of
superconductors in transient magnetic fields is based on the Ginzburg-Landau
free energy. It involves a complex interaction inside the sample
between the network of vortices and defects that usually trap
them; this is analyzed very clearly by Campbell and Evetts \cite{campbell72}. 
The process is difficult to analyze numerically.
Even in 1D, coupling Maxwell's equations with Ginzburg-Landau's
equations for the order parameter to understand how vortices 
are induced and interact in a material is challenging \cite{cdt21}. 
Another point is that temperature effects are ignored; these 
are important because they change the parameters and allow vortex motion.
The final issue is that samples have a size of the order of the cm
while the size of a vortex is about 1 $\mu m$ so that there are about
$10^4$ vortices in a typical sample. For all these reasons, one
should consider a macroscopic model, the critical state model together
with an equation for the temperature.

The critical state model is a phenomenological constitutive equation
$E(J)$ connecting the electric field $E$ to the current density $J$.
This gives rise to a nonlinear diffusion equation for the magnetic field
$B$ coupled to an inhomogeneous diffusion equation for the temperature $T$.
Also note that flux jumps are a dynamical effect and depend on how the external
magnetic field $B_e$ is applied, therefore one needs to specify the
external field dynamics. In addition, many of the flux jumps
described in the literature occur in complicated 2D or even 3D experimental
setups. The modeling is done using a finite element software and few details
are given about the numerical procedure. In particular, it is hard to 
understand the electrodynamics and the mechanisms causing flux jumps. 
To clarify these issues, important questions are \\
what are the time scales of the different phenomena involved in the process? \\
what physical conditions lead to flux jumps? \\
what is the external field dynamics that gives a maximal field trapping and
thus a better magnet?

To answer these questions we considered a 1D situation with a simple
field geometry and a standard constitutive
equation. We wrote the equations in dimensionless form identifying the
typical time scales. Using this model, we examined the influence of
the different parameters in particular the heat capacity. We varied
systematically the pulse magnitude and duration to see how they 
influence the trapped field and the flux jumps. We were able to understand 
the mechanism of flux jumps by detailing the terms in the evolution
of $B$. In addition, we analyzed the trapped flux as a function
of the pulse duration and derived conditions to optimize the
magnet.\\
The article is organized as follows: section 2 presents the physical
configuration and model and its dimensionless form. Section 3 describes
the time dynamics of the magnetic pulse together with the numerical
method used to solve. Sections 4 gives the results for medium
and low temperatures. Finally section 5 discusses the flux jumps
and concludes the article.

\section{Physical model and normalization}

The magnetic behavior of a type II superconductor is given by 
the Maxwell equations and a constitutive law. The Maxwell 
equations read 
\begin{eqnarray}
\label{maxwell-faraday} & \nabla \times \mathbf{E} = -\partial_t \mathbf{B}, \\
\label{maxwell-ampere} & \nabla \times \mathbf{B} = \mu_0 \mathbf{J}, 
\end{eqnarray}
where $\mathbf{E},\mathbf{B}$ are the electric and magnetic fields,
$\mathbf{J}$ is the current density and the term
$\partial_t \mathbf{E}$ is neglected (quasi-static regime).
The  constitutive law is
\begin{equation}
    \mathbf{E} = \mathbf{E}(\mathbf{J}),
\end{equation}
where $\mathbf{E}$ and $\mathbf{J}$ are parallel.
The evolution of the temperature of the sample is given by 
\begin{equation}
\label{thermal}
\rho_d C   \partial_t T = \mathbf{E} \cdot \mathbf{J} + \kappa \Delta T,
\end{equation}
where $\mathbf{E} \cdot \mathbf{J}$ is the power heating per unit volume
due to the magnetic field.

\subsection{Simplified 1D configuration }

To understand in detail the interplay between the magnetic field and the
temperature, we follow the authors of \cite{romero07} and reduce the problem
to one dimension.
\begin{figure}[htb] \label{config1d}
\centerline{
\epsfig{file=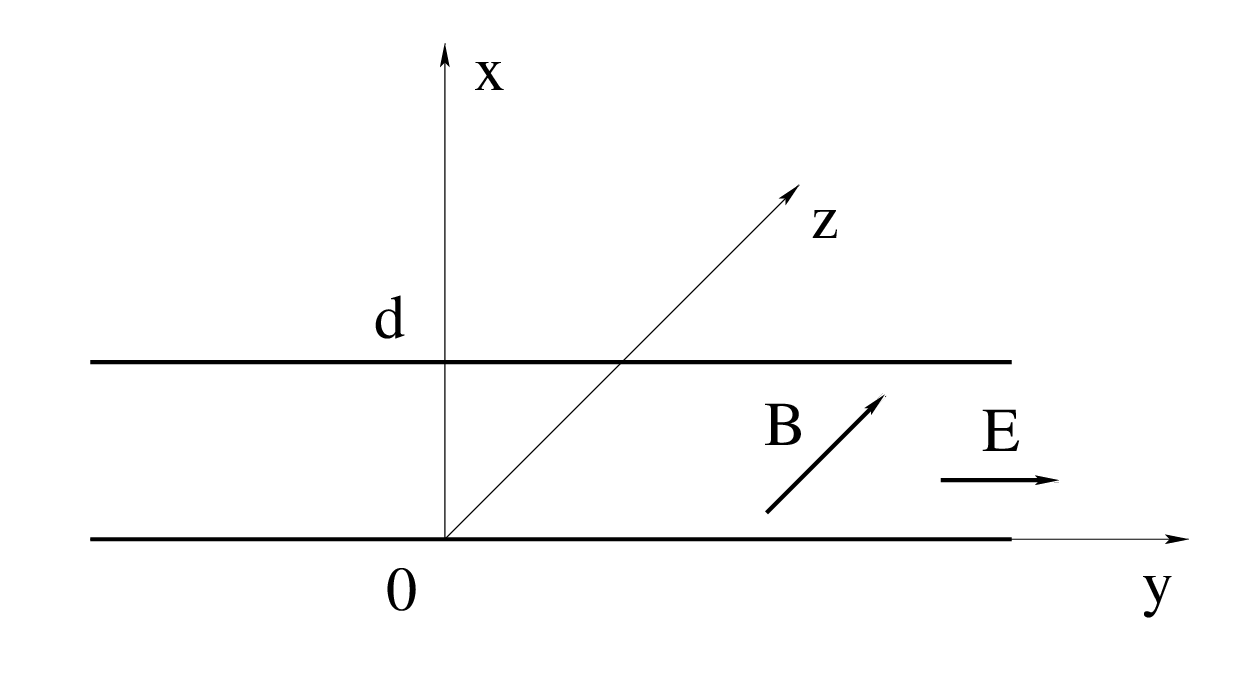, height= 5 cm, width = 12 cm, angle=0}
}
\caption{Simplified configuration: infinite supraconducting plate of thickness $2d$.}
\end{figure}

We consider a
supraconducting plate of thickness $d \gg \lambda$, infinite along $y$ and $z$,
see Fig. \ref{config1d}.
An external time dependent magnetic field $\mathbf{B}_e(t)$ is applied along $z$.
Then we can write 
$$ \mathbf{B} = (0, 0, B(x,t))^T, ~~\mathbf{E} = (0, E(x,t),0)^T  ,~~\mathbf{J} = (0, J(x,t),0)^T $$
so that equations (\ref{maxwell-faraday},\ref{maxwell-ampere}) reduce to
\be\label{bx} {\partial_x}B = -\mu_0 J, \ee
\be\label{bt} {\partial_t}B = -{\partial_x} E.  \ee
{Following \cite{romero07}, the constitutive relation reads}
\be\label{cons1} E = \rho(J), \ee
where {$\rho$ is the antisymmetric function defined by}
\begin{eqnarray}
\label{soj} \rho(J) = 0.5\rho_0 \left [1+\tanh\left({J-J_c\over w}\right) \right ](J-J_c) ,~~J>0\\
\label{soj2} \rho(J) = 0.5\rho_0 \left [1+\tanh\left({J+J_c\over w}\right) \right ](J+J_c) ,~~J<0 .
\end{eqnarray}
where $\rho_0$ is the normal resistivity and $J_c$ the critical current
density.
This form of $\rho(J)$ is a regularized version of the one 
used by Romero-Salazar et al \cite{romero07}, it is plotted in Fig. \ref{fjc}.
It is close to a standard constitutive law
derived from the Bean critical state model \cite{bean}, see also
\cite{cardwell21} for a detailed justification. For 
small $J$, $E=0$ so that there is no heating $E J$. Only for
$J > J_c$, do we have $E>0$ and the heating term $E J$ becomes
significant.
\begin{figure}[H] \label{fjc}
\centerline{
\epsfig{file=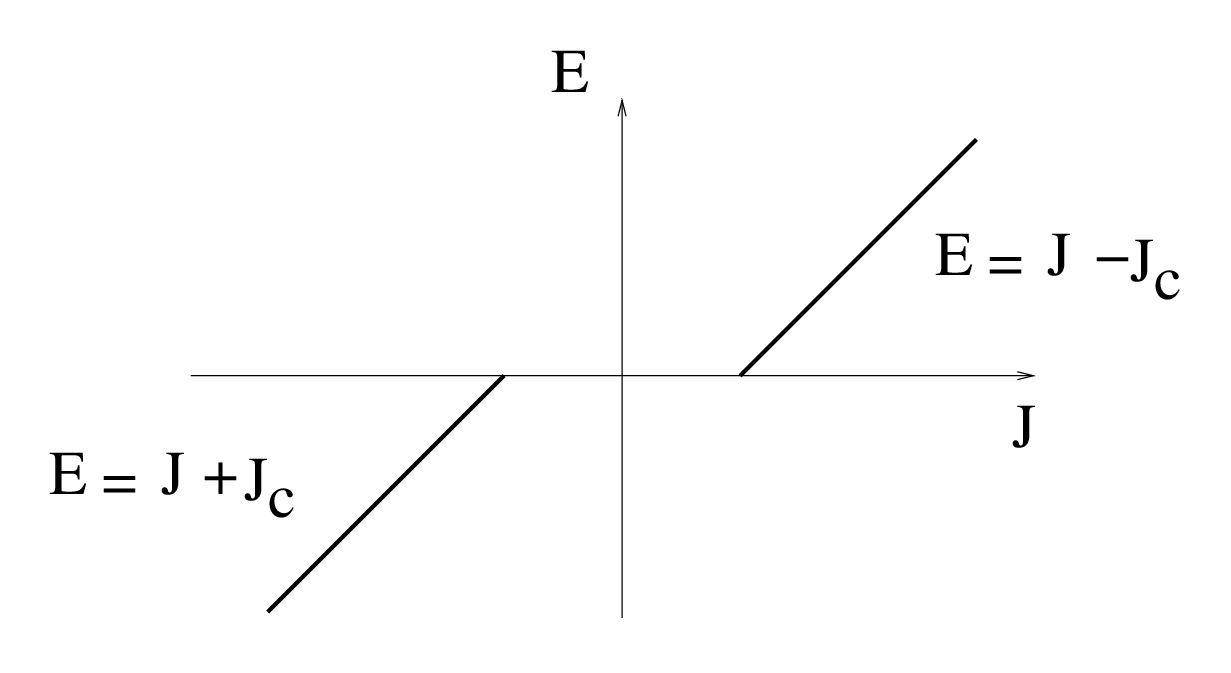, height= 5 cm, width = 12 cm, angle=0}
}
\caption{Plot of $E = \rho(J)$ given by (\ref{cons1},\ref{soj},\ref{soj2}) for small
$w_j$}
\end{figure}
In the following, we will only describe the situation $J>0$ and
$B>0$ for simplicity.
For the critical current density $J_c$, we assume the 
following standard dependence on $T$ and $B$
\begin{eqnarray}
J_c = J_0 \left(1-{T \over T_c}\right)\left(1-{|B| \over B_c}\right)^2~ T< T_c,~|B|< B_c, \label{jc} \\
J_c = 0, ~{\rm otherwise} \nonumber
\end{eqnarray}
where $T_c$ is the critical temperature, $B_c$ is a threshold magnetic field,
and $J_0$ a typical current density. 
Contour plots of $J_c$ are presented in Fig. \ref{jcc} in the plane
$(T,B)$
Note how $J_c$ varies strongly as a function of $B$.
\begin{figure}[H]
\centerline{
\epsfig{file=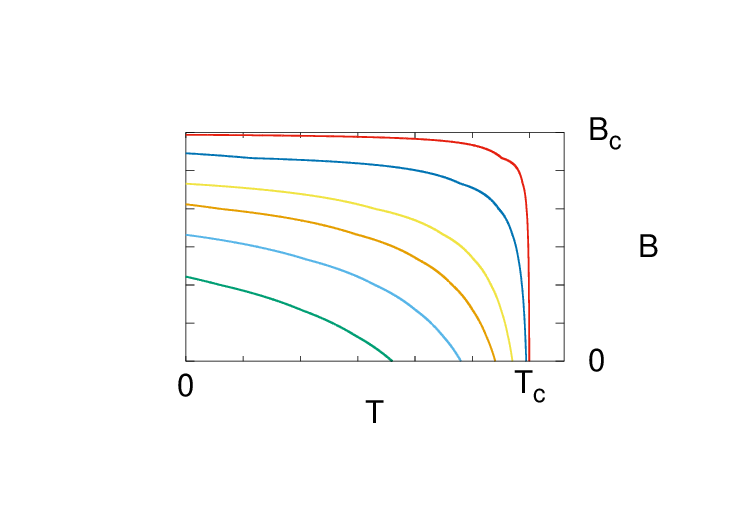, height= 5 cm, width = 12 cm, angle=0}
}
\caption{Plot of $J_c$ given by (\ref{jc}) for
the contour lines 
$0.4,0.2,0.1,0.05,0.01$ 
and $0.001$ from left to right.
}
\label{jcc}
\end{figure}

Collecting equations (\ref{bx},\ref{bt},\ref{cons1}) and the
temperature equation, we obtain the system of equations
\begin{eqnarray}
\label{bt0}{\partial_t} B = -{\partial_x}\left( \rho \left(-{{\partial_x}B \over \mu_0}\right)\right) \\
\label{tt} \rho_m C {\partial_t}T = - {{\partial_x B} \over \mu_0}  \rho \left(-{{\partial_xB} \over \mu_0}\right) + \kappa {\partial^2_{xx}T} .
\end{eqnarray}
on the domain $[0,d]$ together with boundary conditions
\be\label{bc}
B(x=d)=B_e,~~T(x=d)=T_e, ~~\text{symmetry at}~ x=0 .\ee
like in \cite{romero07}.

To gain intuition, assume a simple ohmic behavior $E = \rho_0 J$.
Then the two equations above reduce to diffusion equations and we do not
expect any singular behavior. 
On the other hand, the nonlinearity of the two equations 
(\ref{bt0}, \ref{tt}) gives rise to interesting effects.

\subsection{Time scales }

Table \ref{tpar} summarizes the physical quantities used in the 
model and their dimensions, where $\sM, \sL, \sT, \sI$ and $\sTh$ 
denote the dimensions of mass, length, time, electric current and 
temperature respectively.
\begin{table}[htb]
    \centering
{\small
    \begin{tabular}{|c|c|c|c|}
    \hline
         Symbols & Name & Dimension  & value in SI unit\\
         \hline 
            &             &         &   \\
        $B$ & Magnetic field & $\sM\sT^{-2}\sI^{-1}$ & T\\
        $E$ & Electric field & $\sM\sL\sT^{-3}\sI^{-1}$ & $V m^{-1}$\\
        $J,J_c,J_0,w_J$ & Current density & $\sL^{-2}\sI$ & $A m^{-2}$\\
        $T$ & Temperature & $\sTh$  & K\\ 
            &             &         &   \\
        $x,d$ & Position, width & $\sL$ & m \\
        $t$ & Time & $\sT$ & s\\  
            &             &         &   \\
        $C$ & Heat capacity & $\sL^{2}\sTh^{-1}$ &  $m^{2}~s^2 K^{-1}$\\
        $\rho_m$ & Mass density & $\sM \sL^{-3}$ & $kg~ m^{-3}$ \\
        $\kappa$ & Thermal conductivity & $\sM\sL\sT^{-3}\sTh^{-1}$ & W $m^{-1} K^{-1}$\\ 
            &             &         &   \\
        $\rho_0$ & Resistivity & $\sM\sL^3\sT^{-3}\sI^{-2}$ &  $\Omega$m \\
        $\mu_0$ & Vacuum permeability & $\sM\sL\sT^{-2}\sI^{-2}$ & $kg m s^{-2} A^{-2}$ \\ \hline
    \end{tabular}
}
    \caption{Quantities used in the model, their dimension and unit in the S.I. system. }
    \label{tpar}
\end{table}
In the model, three physical phenomena are coupled: the magnetostatic effects 
described by \eqref{bt0}, the thermal heating due to the electric field 
given by the first term of the right-hand side of \eqref{tt} and
the thermal diffusion described by the second term of the right-hand side of \eqref{tt}.
We can therefore define three characteristic times:
\begin{align}
   &\text{the magnetic timescale:} &\tmag &= \frac{\mu_0d^2}{\rho_0}, \label{tmag}\\
    &\text{the thermal diffusion timescale:}  &\tdif &= \frac{\rho_m C d^2}{\kappa}, \label{tdif}\\
    &\text{the Joule heating timescale:} &\theat &= \frac{\rho_m C T}{EJ} \label{theat}.
\end{align}
To estimate the relative importance of these different time scales,
following \cite{romero07}, we consider a plate of 
thickness $d$ and evaluate the Joule heating term as $E J \simeq \rho_0 J_c^2$.
The values of $d,J_c$ and ${\cal C}= \rho_m C$ are taken from 
\cite[Table 1]{romero07}. They are recalled in Table \ref{tpar2}.
\begin{table}[htb]
    \centering
    \begin{tabular}{|c|c|}
    \hline
         Quantity & value in SI unit\\ \hline
         &            \\
$d$      &   $1$ cm \\
$\rho_0$ &  $10^{-8} \ \Omega$m\\
$\kappa$ & $200$ mW cm$^{-1}$K$^{-1}$ \\ 
${\cal C}= \rho_m C$ &        \\
$\rho_m$ & 2500 kg $m^{-3}$        \\
$C $ &   $50$ $m^{2}~s^2 K^{-1}$    \\
${B_c}$ &  4 T      \\
${T_c}$ &  39K      \\
$J_c$ &  $4~ 10^9 ~A m^{-2}$\\
\hline
    \end{tabular}
    \caption{Parameter values extracted from \cite{romero07}, \cite{fracasso23} and
\cite{Zou2015}. }
    \label{tpar2}
\end{table}
The resistivity $\rho_0$ corresponds to the reciprocal of the normal-state conductivity 
used in \cite{fracasso23} and the thermal conductivity is $\kappa$ as depicted 
in \cite[Fig. 1]{Zou2015} for the lowest temperature ($T=20$K).

The heat capacity of a material like MgB2 depends on the temperature as
shown in Figure of Zou's article \cite{Zou2015}. For 
$20 < T< 50 K$ this dependence
is weak and can be neglected, however when $T \to 0$ $C \to 0$  so that
for small temperatures, one should write
\be\label{coft} C(T) = C_0 T , ~~C_0= 0.25 m^2 s^2  ,\ee
to follow the measurements of \cite{Zou2015}.
This will change the equations and their normalization.

For large temperatures where $C$ can be assumed constant, the resulting time scales are 
\be\label{timesCF}
\tmag \approx 1.25 ~10^{-3}~ s ; ~~ \tdif \approx 0.62 ~s  ; ~~ 
\theat \approx 4.81 ~10^{-4} ~s  . \ee
We therefore conclude that the two dominating effects are 
the magnetic diffusion and the Joule heating, as they occur on a much shorter timescale 
than the thermal diffusion.


\subsection{Normalization }

We can now proceed with the normalization of the equations.
Starting from
\begin{align} \label{sys}
    {\partial_x B} &= -\mu_0 J,\\
    {\partial_tB} &= -{\partial_xE},\\
    E&= \rho(J),\\
    \rho_m C{\partial_tT}&=\kappa{\partial^2_{xx}T}+EJ, 
\end{align}
we introduce
\begin{align*}
x = d x';~t = t_0 t';~ T= T_c T'; ~B= B_c B'\\
E= E_0 E',~J=J_0 J',
\end{align*}
where $E_0$ and $J_0$ will be chosen later and
\be\label{t0}
t_0= \theat ={\rho_m C T_c \over E_0 J_0}, \ee
is a typical Joule heating time $\theat$.
The nonlinear resistivity can be written as
\begin{equation}
    \rho(J) = \rho_0 J_0 \rho'(J'),
\end{equation}
where $\rho_0$ is the normal state resistivity and
\begin{eqnarray}
\label{soj1} \rho'(J') = 0.5 \left [1+\tanh\left({J'-J'_c\over w'}\right) \right ](J'-J'_c) ,\\
\label{jc1} J'_c = (1-T')(1-|B'|)^2,
\end{eqnarray}

\subsubsection{High temperature: $C$ constant}
Plugging the formulas above into equations (\ref{sys}), we
obtain
\begin{align}
    \frac{B_c}{d}{\partial_{x'}B'} &= -\mu_0J_0J',\\
    \frac{B_c}{t_0}{\partial_{t'}B'}&=-\frac{E_0}{d}{\partial_{x'}E'},\\
    E_0 &= \rho_0J_0, \\
    E' &= \rho'(J'), \\
    \frac{\rho_m CT_c}{t_0}{\partial_{t'}T'} &= \frac{\kappa T_c}{d^2}{\partial^2_{x'x'}T'}+E_0J_0E'J'.
\end{align}
It is therefore natural to take the following value for $J_0$
\begin{equation}
    J_0 = {\frac{B_c}{\mu_0 d}}.
\end{equation}

We also define 
\be\label{e0} E_0 = \rho_0 J_0, \ee
to obtain
\begin{align}
\label{erj0}    E &= \rho(J),\\
\label{bx0}    {\partial_xB} &= -J,\\
\label{bt00}    {\partial_tB} &= -\alpha {\partial_xE},\\
\label{tt0}     {\partial_tT} &= EJ + \beta {\partial^2_{xx}T} ,
\end{align}
where the primes have been dropped for clarity and the parameters
$\alpha,\beta $ are given by
\be\label{ab}
\alpha = \frac{t_0}{\tmag}= {\rho_m C T_c \mu_0 \over B_c^2}; ~~
\beta = \frac{t_0}{\tdif} = {\kappa T_c \mu_0^2 \over \rho_0 B_c^2}.\ee
With the values of the parameters, we find 
\be\label{abval1}
\alpha \approx 0.38; ~~\beta \approx 7.69~10^{-4} . \ee

The system of equations (\ref{bx0},\ref{bt00},\ref{erj0},\ref{tt0})
can be written in a compact form by eliminating $E$ and $J$.
We get 
\begin{align}
\label{bt1}    B_t &= \alpha {\partial_x}\left[\rho(B_x)\right],\\
\label{tt1}    T_t &= B_x \rho(B_x) + \beta T_{xx} ,
\end{align}
where the partial derivatives are written as indices to simplify notation.
The minus sign in equations (\ref{bx0},\ref{bt00}) is absent 
since $\rho$ is antisymmetric.

\subsubsection{Low temperature: $C(T)$ }

When $C(T)= C_0 T$, the equations become
\begin{align}
\label{bt2}    B_t &= \alpha {\partial_x}\left[\rho(B_x)\right],\\
\label{tt2}    T T_t &= B_x \rho(B_x) + \beta T_{xx} ,
\end{align}
where the coefficients $\alpha,\beta$ are given by
\be\label{ab1}
\alpha = \frac{t_0}{\tmag}= {\rho_m C_0 T_c^2 \mu_0 \over B_c^2}; ~~
\beta = \frac{t_0}{\tdif} = {\kappa T_c \mu_0^2 \over \rho_0 B_c^2}.\ee
With the values of the parameters, we find
\be\label{abval2}
\alpha \approx 7.5 ~10^{-2} ; ~~\beta \approx 7.69~10^{-4}. \ee
The time scales are now
\be\label{timesCT}
\tmag \approx 1.25 ~10^{-3}~ s ; ~~ \tdif \approx 0.12 ~s  ; ~~
\theat \approx 9.38 ~10^{-5} ~s  . \ee

\subsubsection{Boundary conditions}

From (\ref{bc}) the normalized boundary conditions for both 
(\ref{bt1},\ref{tt1}) and
(\ref{bt2},\ref{tt2})
are
\be\label{bc1}
B(x=1)=B_e,~~T(x=1)=T_e, ~~\text{symmetry at}~ x=0 .\ee
The systems of equations (\ref{bt1},\ref{tt1}) and (\ref{bt2},\ref{tt2})  together with the boundary 
conditions (\ref{bc1}) are our main equations and will be analyzed in the
next sections.

\section{Numerical method}

The system of equations (\ref{bx0},\ref{bt00},\ref{erj0},\ref{tt0})
was discretized in space using second order finite differences and
the time advance was done using an ordinary differential equation solver,
typically a Runge-Kutta method.

For the spatial discretization, we introduce an
$N$-points grid for the spatial domain $[0,1]$
\begin{equation}
0=x_0 < x_1 < \dots < x_{N-1}  = 1,
\end{equation}
where
\begin{equation}
x_j = j\Deltx, \quad\quad \text{ and } \quad\quad \Deltx = \frac{1}{N-1}.
\end{equation}
Let $B_j^n$, $T_j^n$, $E_j^n$ and $J_j^n$ be the approximations of $B$, $T$, $E$ and $J$ at time 
$t_n = n \Deltt$ and at point $x_j$ respectively.
For constant $C$, assuming a simple Euler time discretization, equations 
(\ref{bx0},\ref{bt00},\ref{erj0},\ref{tt0}) yield
\begin{align}
\label{a1} J_j^n & = \frac{B^n_{j+1}-B^n_{j-1}}{2\Deltx},\\
\label{a2} E_j^n & = \rho(J_j^n,T_j^n), \\
\label{a3} {B_j^{n+1}- B_j^n \over \Delta t} & = \alpha {E_{j+1}^{n}- E_{j-1}^n \over 2 \Delta x}, \\
\label{a4} {T_j^{n+1}- T_j^n \over \Delta t} & = E_j^nJ_j^n + \beta \frac{T_{j+1}-2T_j^n+T_{j-1}^n}{\Deltx^2} , 
\end{align}
for interior grid points. For variable $C$, the last equation reads
$$T_j^{n}({T_j^{n+1}- T_j^n \over \Delta t})  = E_j^nJ_j^n + \beta \frac{T_{j+1}-2T_j^n+T_{j-1}^n}{\Deltx^2}.$$ 
We used 
off-centered differences to handle the two boundary points
$x=0,~1$. The time advance from $B_j^n,~ T_j^n$ to $B_j^{n+1},~ T_j^{n+1}$ uses the
following algorithm
\begin{itemize}
\item[(i)] Compute $J_j^n$ using (\ref{a1}),
\item[(ii)] Compute $E_j^n$ using (\ref{a2}),
\item[(iii)] Compute  $B_j^{n+1}, T_j^{n+1}$ using (\ref{a3},\ref{a4}).
\end{itemize}

To ensure the stability of the method, $N$ or $\Deltt$ are chosen such 
that the Courant-Friedrich-Levy stability condition \cite{strauss}
\begin{equation}
\alpha \frac{\Deltt}{\Deltx^2}< 1,
\end{equation}
holds. 
To gain accuracy, we used an explicit fourth-order Runge-Kutta method 
for the time integration. In most runs, we used
$$ \Deltx = 10^{-2}, ~~ \Deltt = 10^{-4},$$
because $\alpha$ is small. 
{Some numerical experiments have also been carried out using a \emph{Total Variation Diminishing} (TVD) Runge-Kutta scheme, leading to similar numerical results.}

\subsection{Magnetic field pulse}

Before describing in detail the results, we make the following remarks.
\begin{itemize}
\itemsep -0.1 cm
\item Care must be taken in handling the situations $E<0$ and $J<0$. 
\item We observe numerically that calculations cannot be started with
$B_e$ large because of instabilities.
\item Because of the symmetry condition, $B(x)$ is not differentiable
at $x=0$. We therefore enforce $J(0)= \partial_x(B) (0)=0$ to stabilize 
the numerical method. Otherwise, we observe an instability of the scheme.
\item The critical current $J_c$ is temperature dependent and it is zero
for $T> T_c$.
\end{itemize}

We use a magnetic field $B_e(t)$ that ramps up from 0 to $B_{max}$ 
and back to 0,  see Fig. \ref{bump09be}.
\begin{figure}[H]
\centerline{
\epsfig{file=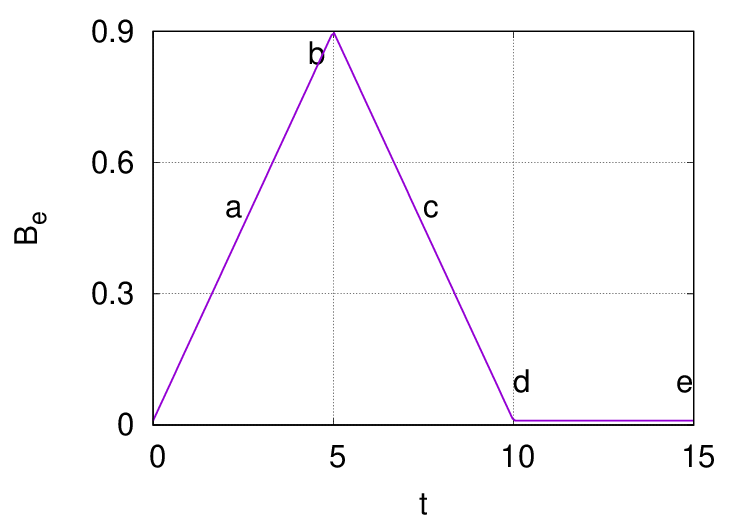, height= 5 cm, width = 10 cm, angle=0}
}
\caption{Plot of $B_e(t)$ for a magnetic field pulse with $B_{max}=0.9$.}
\label{bump09be}
\end{figure}
We denote by $t_p$ the duration of the magnetic pulse $B_e(t)$,  in Fig. \ref{bump09be}  $t_p = 10$.

\section{Numerical results}

\subsection{High temperature (constant $C$)}

We first consider a medium temperature $T < 1 (=T_c)$ for which 
we can assume the heat capacity to be constant. Then we use 
equations (\ref{bt1},\ref{tt1}) and parameters $\alpha,\beta$ 
given by (\ref{abval1}).


We first choose $B_{max}=0.5$ and $T_e=0.5$ and a pulse duration
$t_p=10$.
Fig. \ref{cfT05B05} shows snapshots of $B(x), T(x), E(x)$ and 
$J(x)$ from top left to bottom right for the five time instants
a, b, c, d and e corresponding to the magnetic pulse of Fig. 
\ref{bump09be}. Note how B increases gradually. 
\begin{figure}[H]
\centerline{
\epsfig{file=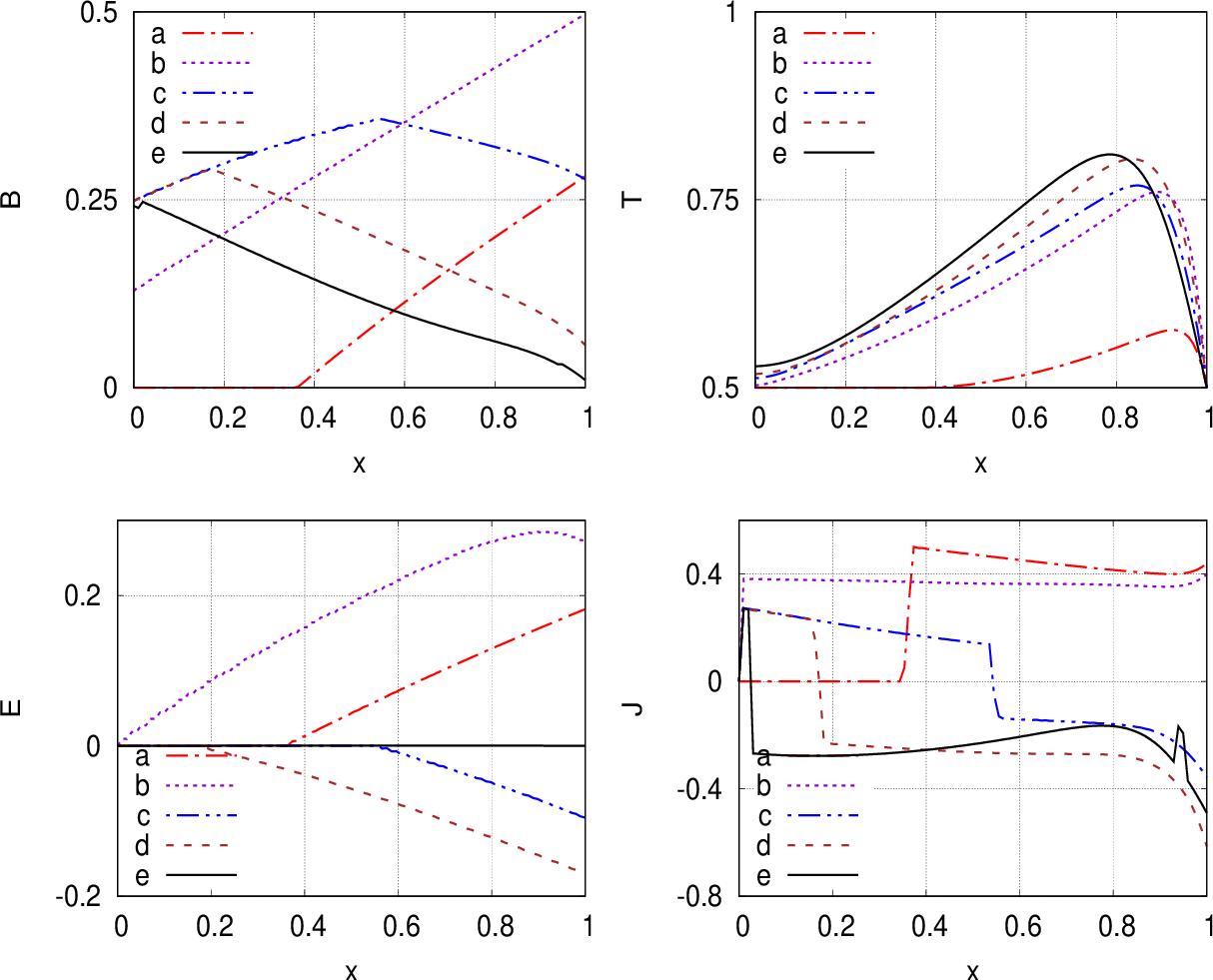, height= 12 cm, width = 7 cm, angle=90}
}
\caption{Snapshots of $B(x),T(x)$ (top) and $E(x),J(x)$ (bottom) for
a field pulse with $B_{max}=0.5$ and $t_p=10$. The plots correspond 
respectively to labels a,b,c,d and e in Fig. \ref{bump09be}.
The external temperature is $T_e=0.5$.}
\label{cfT05B05}
\end{figure}
For snapshot (b) $B$ is blocked at $x=0.4$ 
because $J=0$ for $x >0.4$; there is no diffusion.
At (c) we have a large temperature increase so that $B$
increases strongly. The temperature does not change significantly
after that. For snapshot (e), after the pulse has passed, there 
is a trapped field $B$ and a non zero current $J$.

Increasing $B_{max}$ to 0.9, we observe that the sample becomes
normal. The snapshots are shown in Fig. \ref{cfT05B09}.
\begin{figure}[H]
\centerline{
\epsfig{file=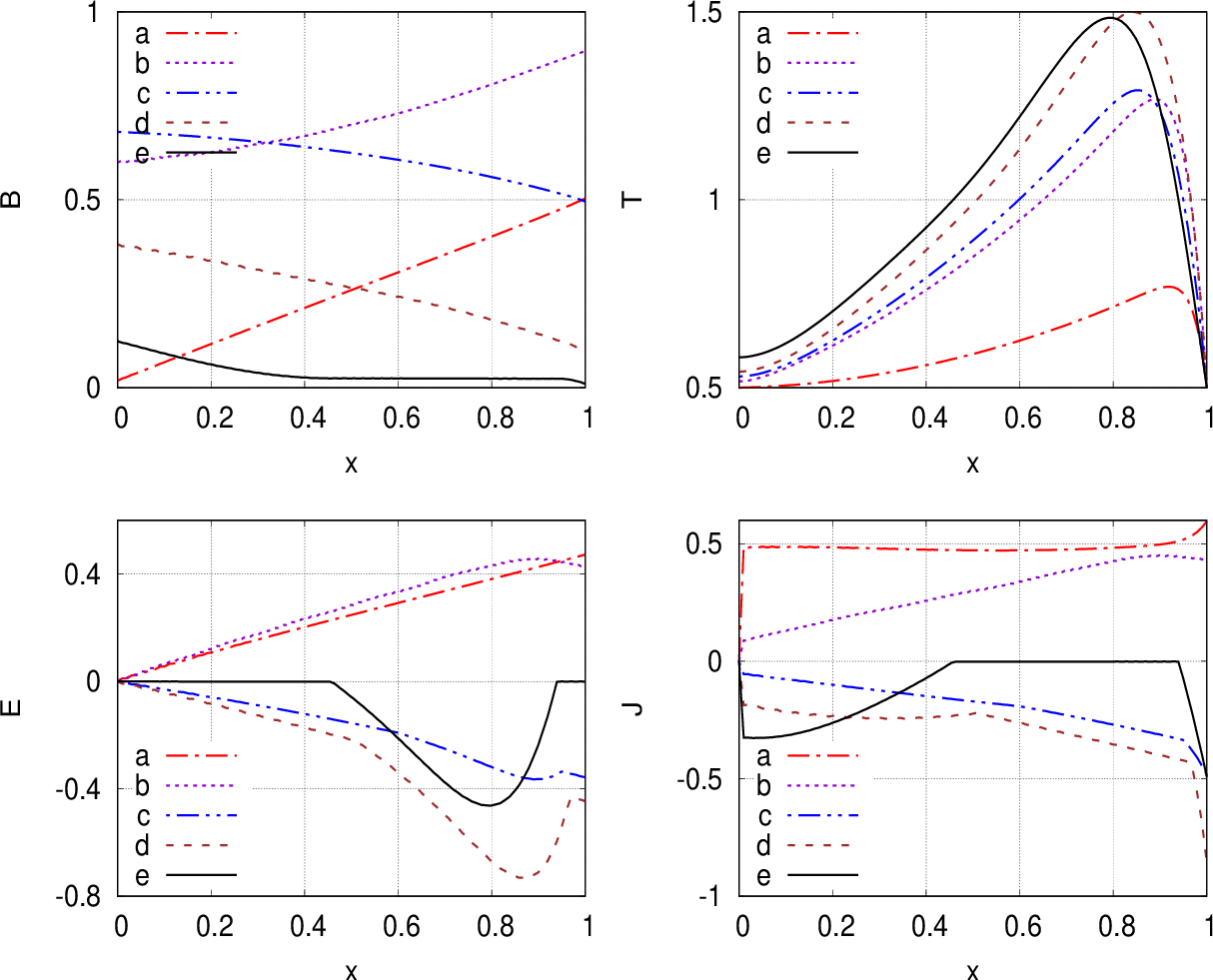, height= 12 cm, width = 7 cm, angle=90}
}
\caption{Same as Fig. \ref{cfT05B05} except $B_{max}=0.9$.}
\label{cfT05B09}
\end{figure}
Notice the linear profile for (a), in agreement with (\ref{b0}).
The temperature develops a gradient so that the last term in
(\ref{Bt}) becomes significant and causes a large increase in $B$
for snapshot (b). After the pulse has passed, the trapped field is small
even though $B_{max}$ was large. Notice the zero regions in $E$
and $J$ where $J_c=0$.

\subsubsection{Magnetization}

To measure the trapped field induced by a magnetic field pulse,
we computed the magnetization in the sample as
\be\label{mag}
M = \int_0^1 (B(x) -B_e) dx \ee
\begin{figure}[H]
\centerline{ \epsfig{file=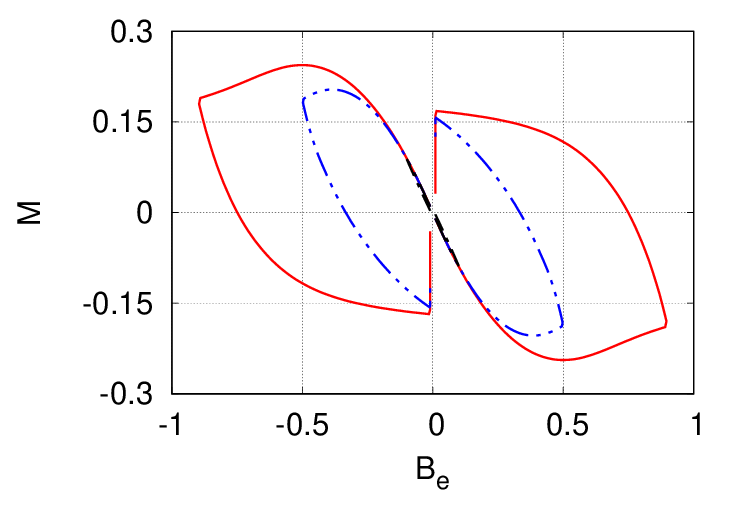, height= 5 cm, width = 12 cm, angle=0} }
\caption{Plots of $M(B_e)$ for field pulses of duration $t_p=10$ and
$B_{max}=0.9$ (continuous curve, red online), $0.5$ (long dash, blue online) 
and $0.1$ (short dash, black online).
The external temperature is $T_e=0.5$.}
\label{cfmbe}
\end{figure}
The curve $M(B_e)$ for $B_e$ ramped up to $B_{max}=0.9$ (red online) 
drops down to 0 after the pulse has passed. This is because
the sample has turned normal so that there is no trapped field.
On the contrary for $B_{max}=0.5$ (blue online), there
is a small dip that stops. Here, the sample remains superconductor
and there is a trapped field. For $B_{max}=0.1$ (black online) the
trapped field is very small, this shows that nonlinear effects are 
important to trap a magnetic field.
Finally note that the curves are smooth, so that there are no 
flux jumps for these parameters.

To prevent the sample from turning normal, one can lengthen the 
duration $t_p$ of the magnetic field pulse. Taking the parameters of
Fig. \ref{cfT05B09} and applying the pulse over a duration $t_p= 100$ 
instead of 10 prevents the sample from turning normal. The results are
presented in Fig. \ref{cft100T05B09}.
Again we observe a linear profile for snapshot (a) in agreement with 
(\ref{b0},\ref{bk}). The temperature does not change much and $B$ follows
$B_e$. Notice the trapped field after the pulse has passed, it is comparable
to the one for $B_{max}=0.5$nd $t_p=10$.
\begin{figure}[H]
\centerline{
\epsfig{file=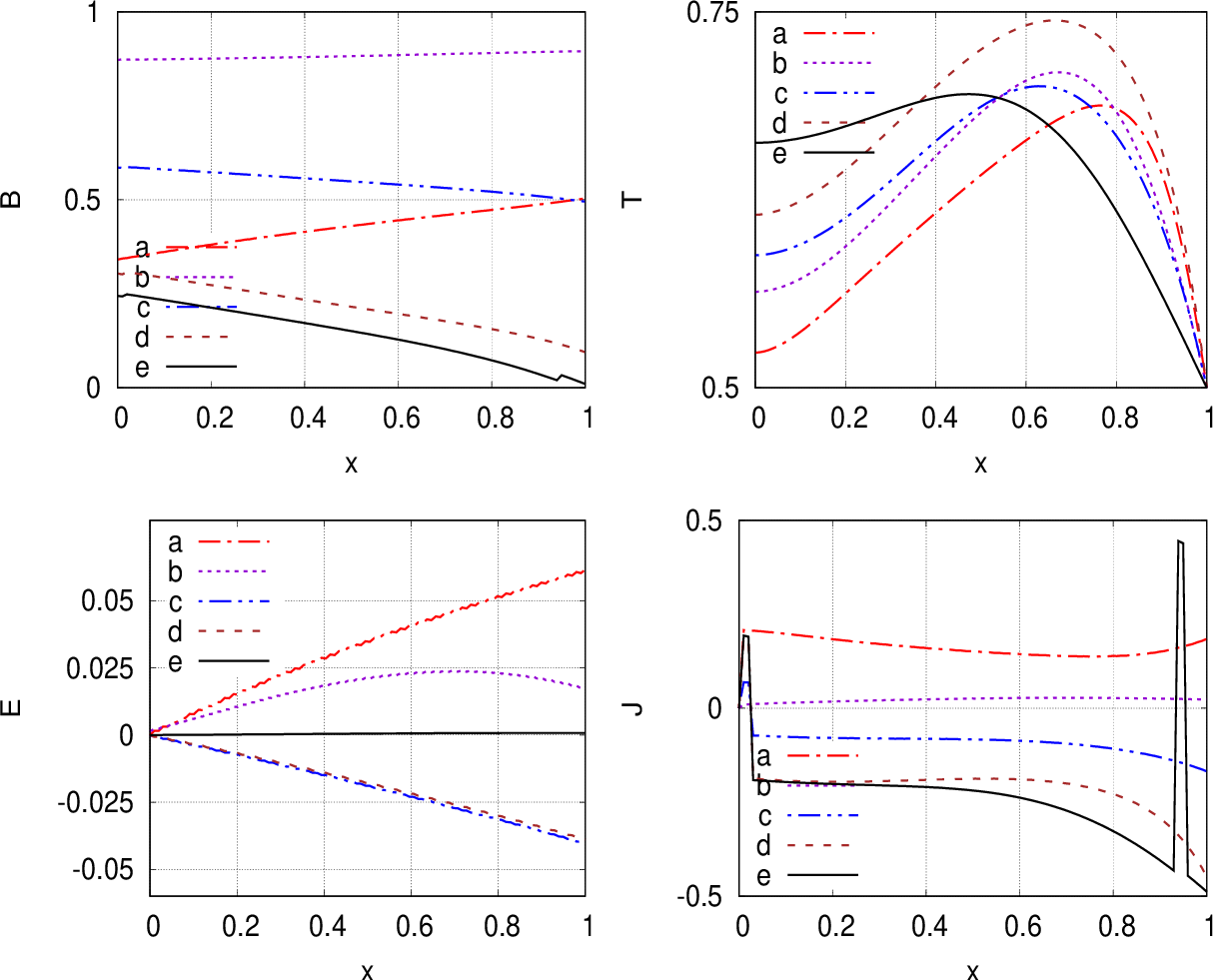, height= 12 cm, width = 7 cm, angle=90}
}
\caption{Same as Fig. \ref{cfT05B09} except $B_e$ is ramped up
over a duration $t_p= 100$.}
\label{cft100T05B09}
\end{figure}

\begin{figure}[H]
\centerline{ \epsfig{file=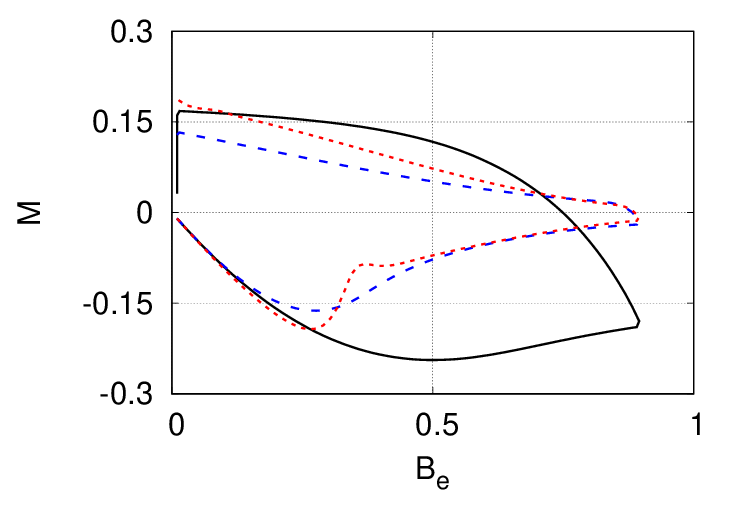, height= 5 cm, width = 12 cm, angle=0} }
\caption{Magnetization for pulse duration $t_p= 10$ (black online)
100 (long dash, blue online) and 1000 (short dash, red online).}
\label{TIMEcfT05mbe}
\end{figure}
The magnetization is shown in Fig. \ref{TIMEcfT05mbe} for 
$t_p=10,~100$ and 1000.  
For a duration $t_p= 10$, there is no trapped field as the sample has turned
normal. Notice how $M$ decays slowly for the red curve. For a duration
of 100, the sample remains superconductor, there is a trapped field.
The trapped field is larger for a duration of 1000 since the heating caused
by the external field has time to dissipate.

\subsection{Low temperature ($C(T)$) and flux jumps }

For small temperatures, the dependence $C(T)$ cannot be neglected.
We now study solutions of the system of equations
(\ref{bt2},\ref{tt2}) and parameters $\alpha,\beta$
given by (\ref{abval2}). Throughout this section, we assume
a small temperature $T_e=0.1$.
We first consider a pulse $B_{max}=0.5$ of duration $t_p=10$ as in the previous
section. The snapshots are shown in Fig. \ref{ctT01B05}.
Notice the large temperature increase from (a) to (b), there part
of the sample has lost its superconductivity. Then $J_c=0$ and $B$
is small for snapshot (e) and will continue to decay. Notice how both
$E$ and $J$ are not zero for (e) so the term $E J$ remains significant
and drives this decay of B.
\begin{figure}[H]
\centerline{
\epsfig{file=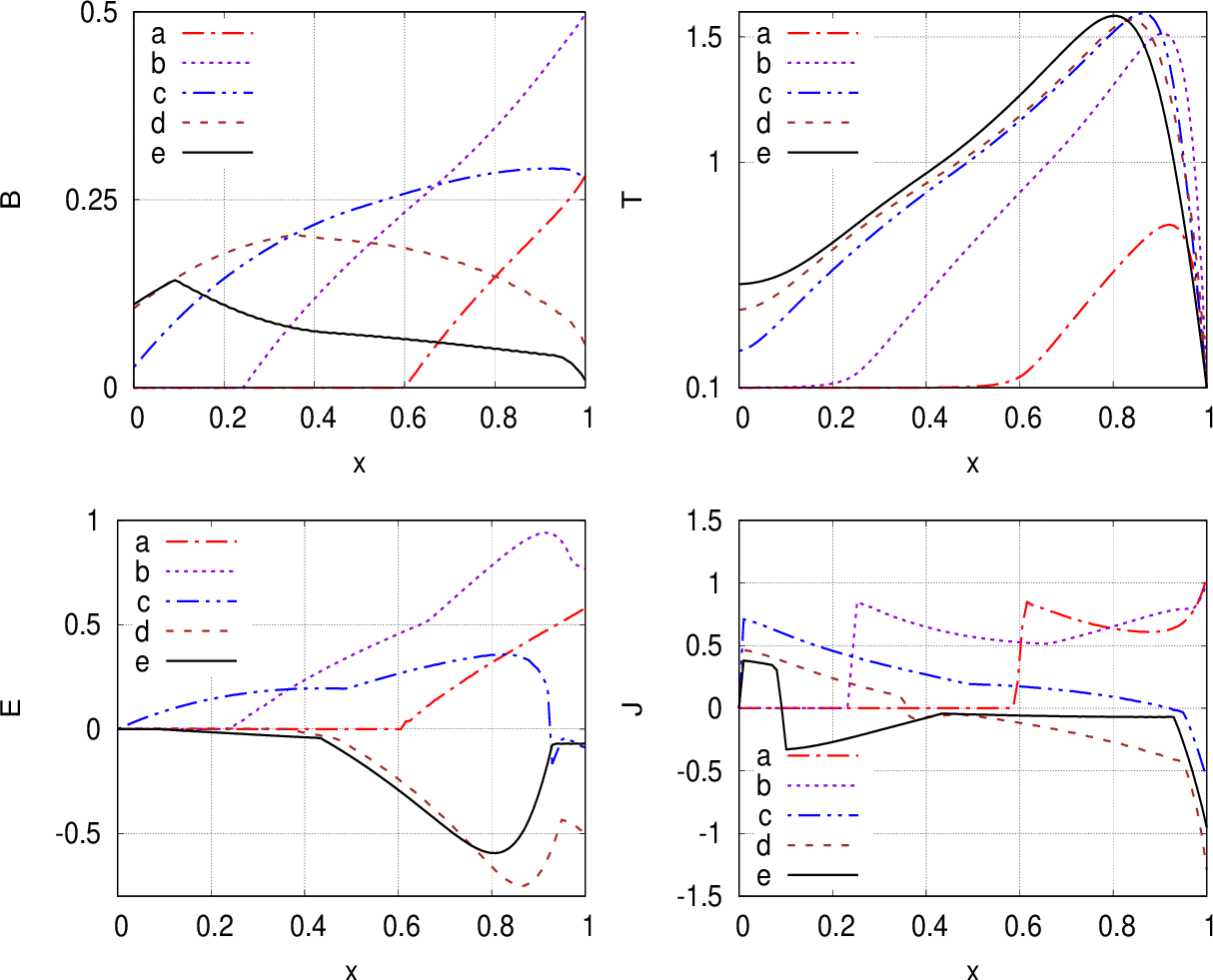, height= 12 cm, width = 7 cm, angle=90}
}
\caption{Snapshots of $B(x),T(x)$ (top) and $E(x),J(x)$ (bottom) for
a field pulse with $B_{max}=0.5$ and $t_p=10$ for.
labels a,b,c,d and e in Fig. \ref{bump09be}.}
\label{ctT01B05}
\end{figure}

Increasing the duration of the pulse to $t_p= 100$ for the 
last set of parameters does not prevent the sample from turning normal. 
This is probably due to the temperature equation which evolves $T^2$.
Only increasing to
a duration  $t_p = 1000$ does the sample remain superconductor.

\subsubsection{Flux jump in magnetization }

We observed a small flux jump for a long duration pulse for a
constant heat capacity $C$. Let us examine the situation $C(T)$.
Fig. \ref{ctT01topmbe} shows the magnetization vs $B_e$ for
pulse durations $t_p= 600, 1500, 3000$. Notice that $M(B_e)$ is smooth
for $t_p=600$ and presents large large flux jumps for $t_p= 1500$ and 3000.
The position of the flux jumps depends on the duration of the pulse, more
precisely on the slope ${d B_e \over dt}$. For $B_{max}=0.5$, we tested
different times of increase of $B_e$ from 0 to $B_{max}$: 1000 , 2000 and
3000 and observed flux jumps at different positions. Also the time 
duration for which $B_e = B_{max}$ does not affect the position
of the flux jumps. 
\begin{figure}[H]
\centerline{ \epsfig{file=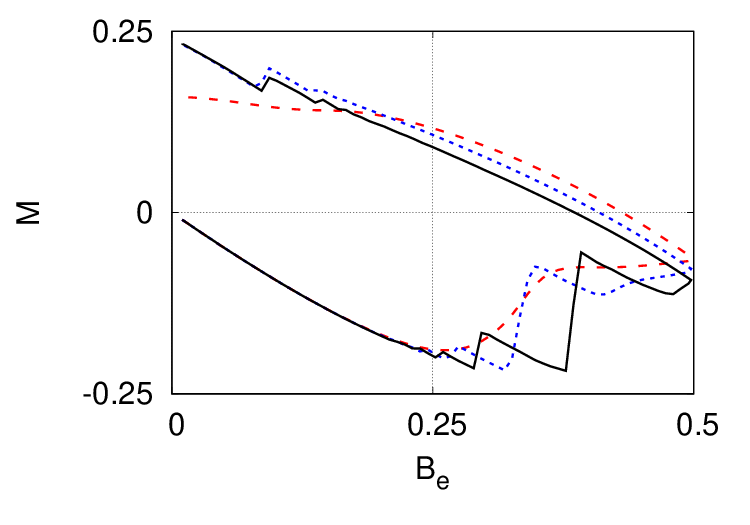, height= 5 cm, width = 12 cm, angle=0} }
\caption{Magnetization $M(B_e)$ for field pulses with $B_{max}=0.5$ 
and pulse durations $t_p= 600$ 
(long dash, red online), 
1500 (short dash, blue online) and 3000 (continuous, black online). The temperature is $T_e=0.1$.}
\label{ctT01topmbe}
\end{figure}

To understand the mechanism of these flux jumps, we consider in detail
the sample with a pulse duration $t_p=3000$. The flux jumps occur for
$B_e=0.3$ and 0.37 as $B_e$ is increasing. The snapshots are
shown in  Fig.  \ref{ctT01B09tp3000}. As expected there is
a large increase of temperature from (a) to (b). The field
$B$ also exhibits a large increase so that $M$ is reduced. Notice how $E J$ is
negative for snapshots (b) and (d). 
\begin{figure}[H]
\centerline{
\epsfig{file=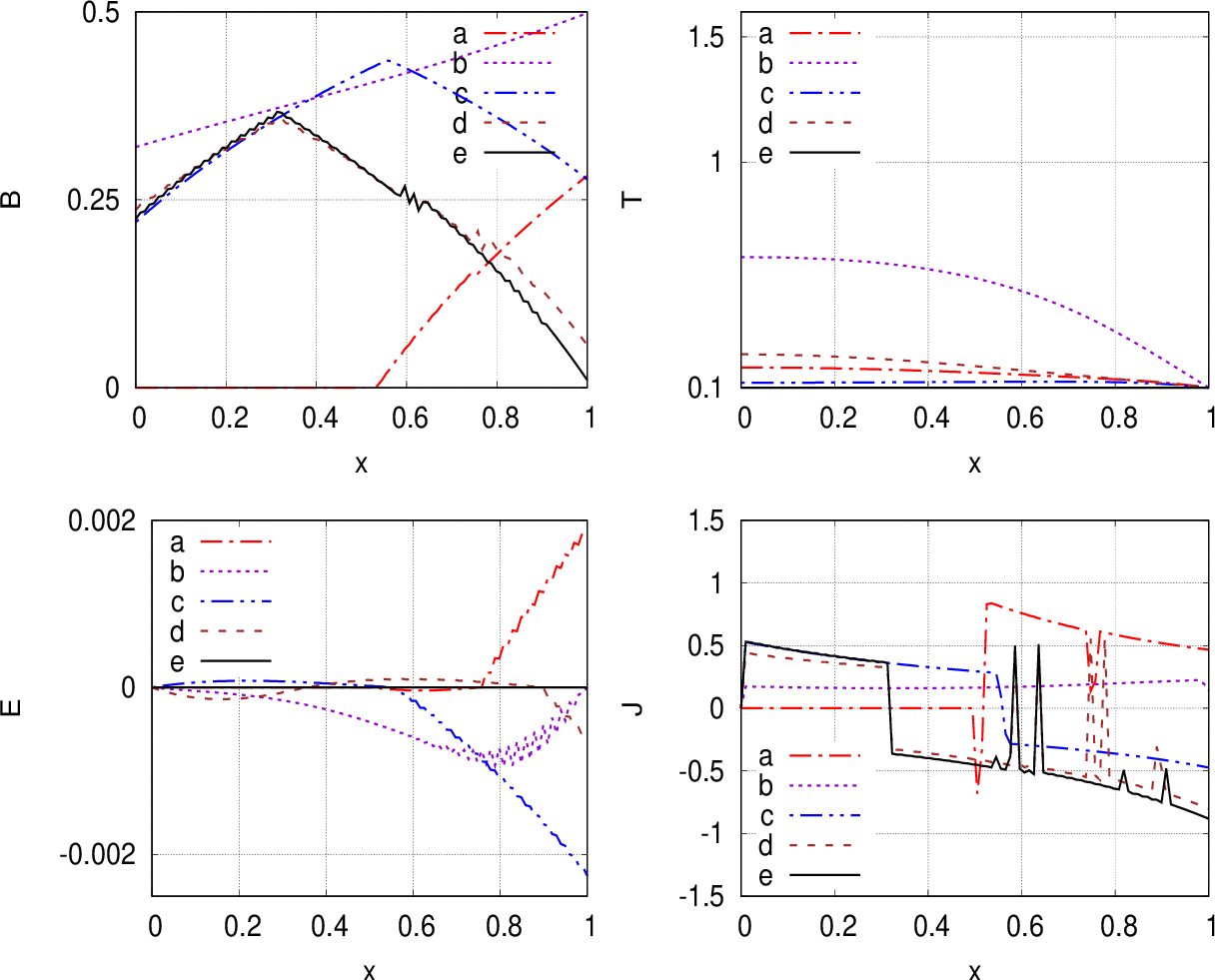, height= 12 cm, width = 7 cm, angle=90}
}
\caption{Plots of $B(x),T(x)$ (top) and $E(x),J(x)$ (bottom) for
a field pulse with $B_{max}=0.5$ and $t_p=3000$.
The external temperature is $T_e=0.1$.}
\label{ctT01B09tp3000}
\end{figure}

We now present a detailed analysis of the two flux jumps observed for
$B_e=0.3$ and $B_e=0.37$. The fields $B(x),T(x),E(x),J(x)$ and $J-J_c(x)$
are shown in Fig. \ref{fj38ct3000} for times $t=570, 585$ and $600$. 
The time step is $\Delta t = 10^{-4}$ in units of $t_{heat}$. Note 
how $E$ and $J-J_c$ are large for $t=585$. 
The temperature increases strongly from $t=570$ to $t=585$ so the third
term in equation (\ref{Bt2}) becomes large and causes a large increase in
$B$ observed for $t=585$. Notice how $B$ does not vary much after 
the flux jump from $t=585$ to 600. In fact, $J$ very close to $J_c$
throughout the increase of $B_e$, except at the flux jumps.
The oscillations of $E$ are due to $J \approx J_c$. The spikes observed
in $J$ correspond to jumps in $B_x$, these are due to the irregular
Burger's front dynamics of $B$.

\begin{figure}[H]
\centerline{
\epsfig{file=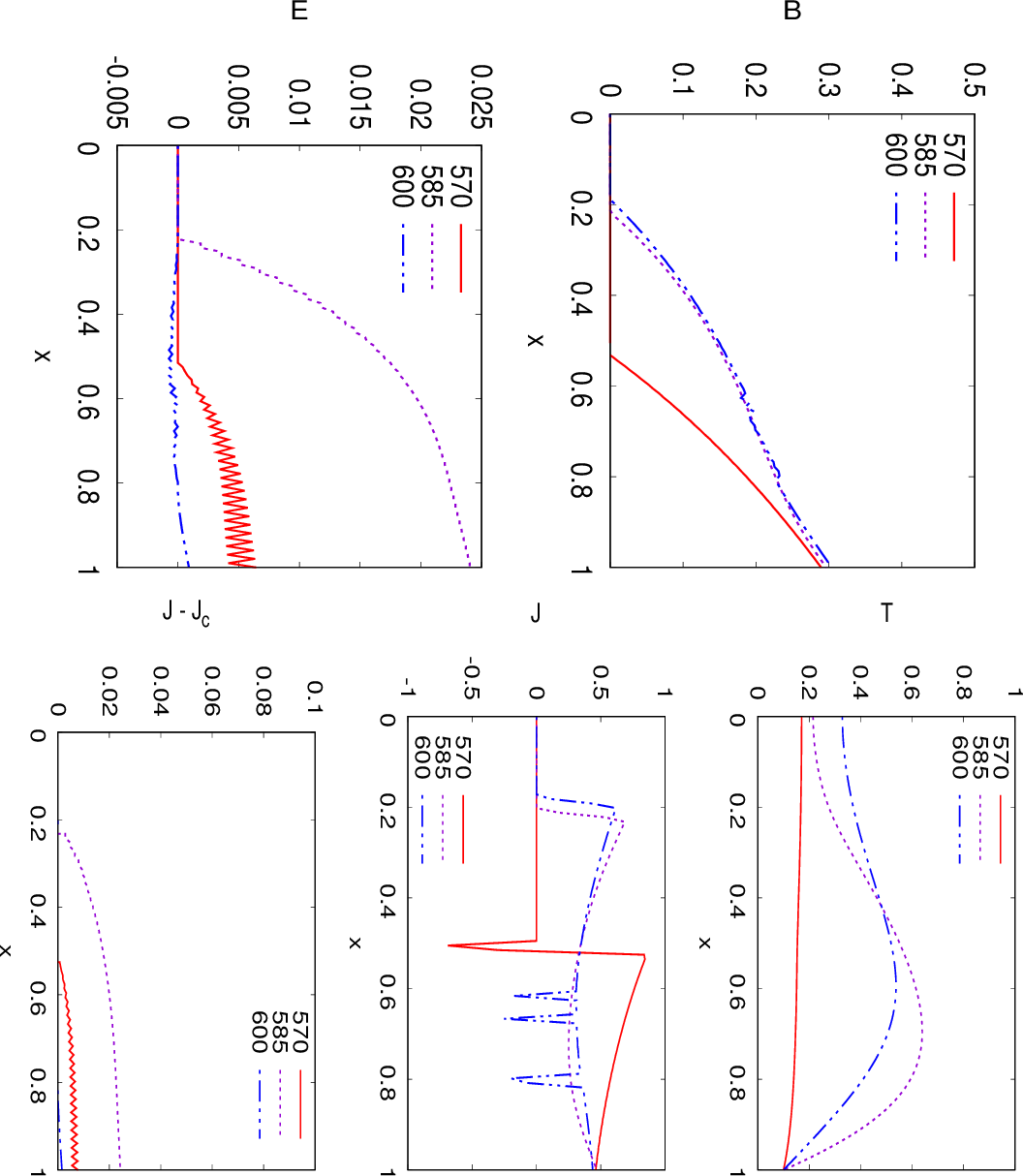, height= 12 cm, width = 7 cm, angle=90}
}
\caption{Plots of $B(x),T(x)$ (top) and $E(x),J(x),J_c(x)$ (bottom) for
three successive times $t=570, 585$ and $600$. 
}
\label{fj38ct3000}
\end{figure}

The dynamics of the other large flux jump observed for $B_e=0.37$
is shown in Fig. \ref{fj50ct3000}. Notice the large increase in $T$
from $t=750$ to 765 so that $B$ has increased. The
electric field is large for $t=765$ so that the $E J$ term
causes a large increase in temperature. This quantity
continues to evolve at $t=780$  so that $B$ increases again. At time
$t=795$, $B$ has settled down, there is little difference between
$B(780)$ and $B(795)$. The temperature continues to evolve however
because of the $B_x \rho(B_x)$ term.
\begin{figure}[H]
\centerline{
\epsfig{file=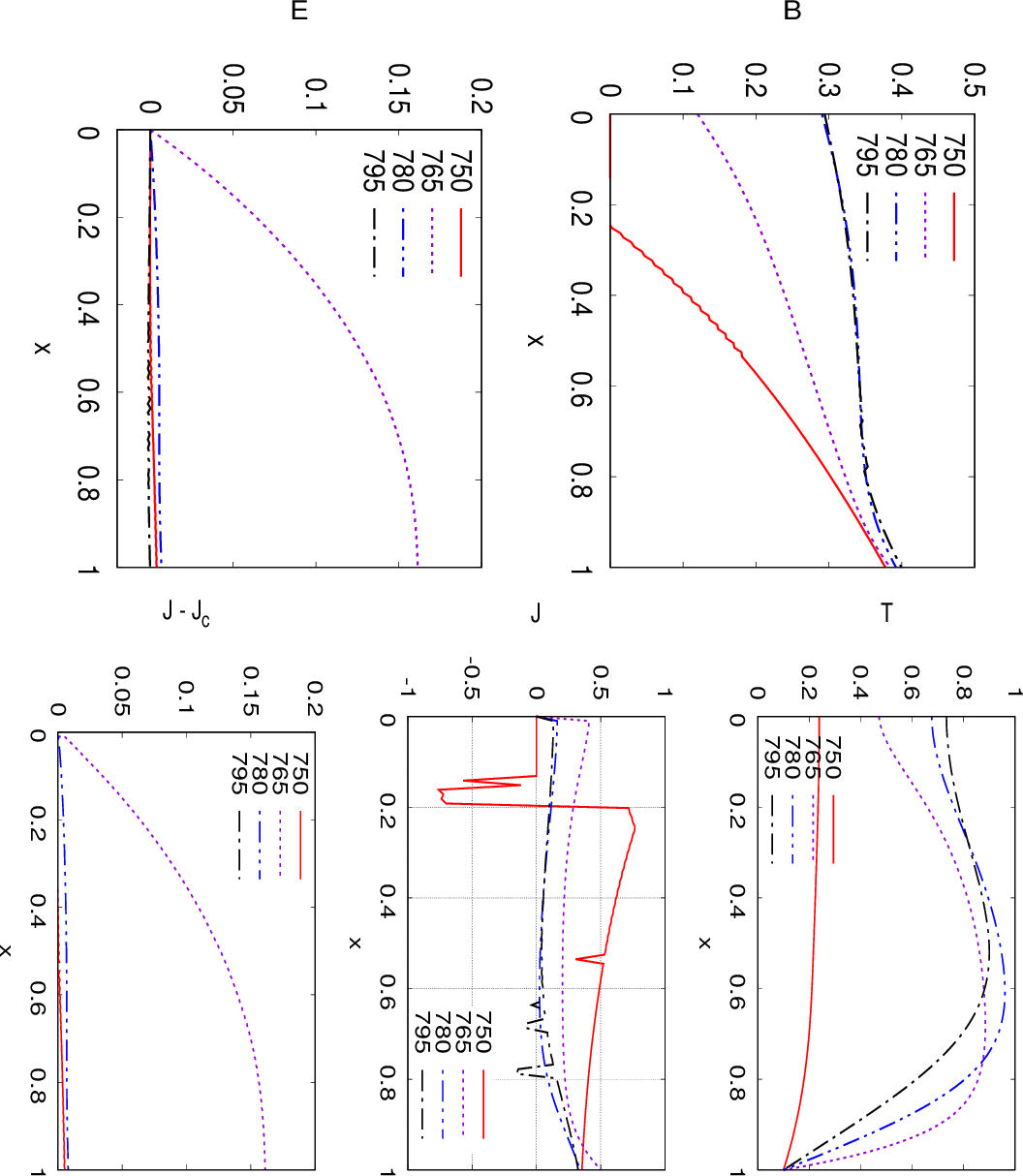, height= 12 cm, width = 7 cm, angle=90}
}
\caption{Plots of $B(x),T(x)$ (top) and $E(x),J(x),J-J_c(x)$ (bottom) for
three successive times $t=750, 765, 780$ and $795$.
}
\label{fj50ct3000}
\end{figure}

\section{Discussion and conclusion}

The results presented above can be understood at least qualitatively by examining the fixed points of the systems of equations
(\ref{bt1},\ref{tt1}) and (\ref{bt2},\ref{tt2}). This analysis
is relevant for magnetic pulses verifying $t_p > t_{mag}$ so that 
both $B$ and $T$ fields reach their fixed point
at each instant of time.
For simplicity, consider the system (\ref{bt1},\ref{tt1})
\begin{align}
\label{bt1a}    B_t &= \alpha {\partial_x}\left[\rho(B_x)\right],\\
\label{tt1a}    T_t &= B_x \rho(B_x) + \beta T_{xx} ,
\end{align}
the analysis will be similar for (\ref{bt2},\ref{tt2}).
The fixed points of this system verify
\be \label{fix} \rho(B_x) = C,~~~ C B + \beta T_{x} = D ,\ee
where $C,D$ are constants.
An obvious fixed point is 
\be \label{fix0} \rho(B_x) = 0,~ T=T_e .\ee
The flux jumps correspond to the rapid switch of the system from
the fixed point $C>0$ to the one where $C=0$.

Examining $\rho(B_x)$ from Fig. \ref{fjc}, one sees that $B_x < J_c$ 
is not interesting because there is no motion of $B$. This is the Meissner
state where the external field $B_e$ is perfectly screened by the 
superconductor. Only for $B_x \approx J_c$ do we have motion, this is the
flux creep state analyzed by Mints \cite{mints96}, see also the
interesting discussion by Moseley et al \cite{moseley21}.
For the fixed point $\rho(B_x) =0$
\be \label{fix1a} B_x = J_c = (1-T)(1-B)^2, \ee
the solution $B(x)$ is given by
\be\label{b0}
B_0(x) = 1 - {1 \over (1-T)(x-1) + (1-B_e)^{-1}} , \ee
see details in the appendix.
For $\rho(B_x) >0$, note that $\rho(B_x) = B_x - J_c$ so that the
second type of fixed point can be obtained by writing
$$B_x = J_c +C = (1-T)(1-B)^2 + C , $$
leading to 
\be\label{bk}
B_C(x) = 1 - \sqrt{C \over 1-T} \tan^{-1}\left [ \sqrt{C (1-T)}(x-1)+ {\rm atan}(\sqrt{C \over 1-T} {1 \over 1-B_e} ) \right ]
. \ee
As expected $B_C(x) \to B_0(x)$ for $C \to 0$.
Note that this is an approximation because we neglect the dependance $T(x)$.
In fact, it is not clear that there exists a solution to the system
(\ref{fix}).
Also remark that the model $E= \rho(J)$ shown in (\ref{fjc}) and used throughout the
article allows to have solutions where $J > J_c$ as opposed to the
commonly used model $E= (J/J_c)^{n}, ~~n \approx 40$ for which 
$J>J_c$ is not possible, see the comments in \cite{moseley21}.

Fig. \ref{compbxjc} shows the two fixed points $B_0(x)$ and $B_C(x)$ 
for $C=0.1, ~T=0.5$ and $B_e=0.195, ~0.27$ and $0.45$.
\begin{figure}[H]
\centerline{
\epsfig{file=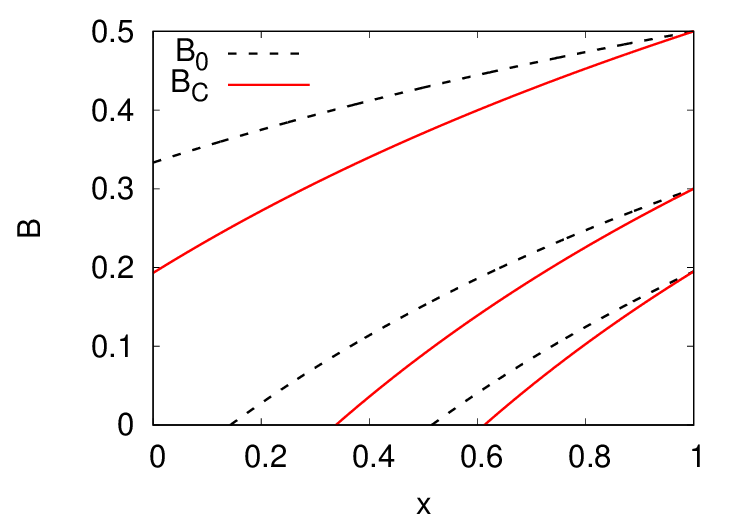, height= 5 cm, width = 12 cm, angle=0}
}
\caption{ The two fixed points $B_0(x)$ and $B_C(x)$ for
 $B_e=0.195, ~0.27$ and $0.45$.}
\label{compbxjc}
\end{figure}
To illustrate the importance of these fixed points in the dynamics 
we show in Fig. \ref{bxjc} the inumerical solution 
$B(x,t)$ for $B_e=0.195, ~0.27$ and $0.45$ (dashed lines)
together with
the two fixed points $B_0(x)$ (left panels) and $B_C(x), ~C=0.1$ (right panels)
drawn in continuous line (red online).
The top panels correspond 
to a pulse with $B_{max}=0.5,~ t_p = 10, T_e=0.5$ and the bottom panels
to a pulse with $B_{max}=0.5,~ t_p = 1000, ~ T_e=0.1$ for 
$B_e=0.195, 0.27$ and $0.45$.
For the top panels the data fits better $B_C(x)$ while for the
bottom panels the data fits better $B_0(x)$.
\begin{figure}[H]
\centerline{
\epsfig{file=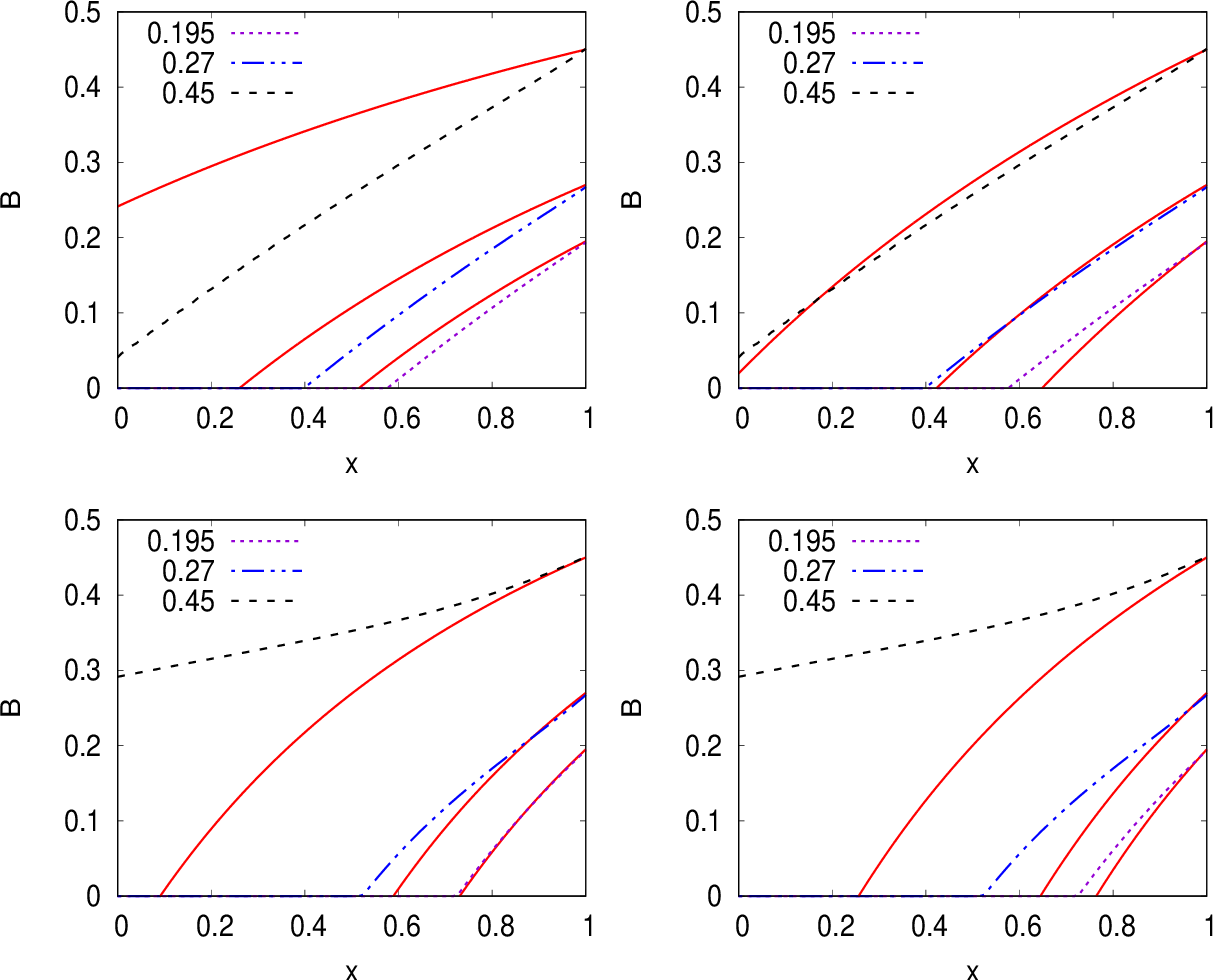, height= 12 cm, width = 5 cm, angle=90}
}
\caption{Comparison of the two fixed points $B_0(x)$ and $B_C(x)$ with
	numerical results. See text for details}
\label{bxjc}
\end{figure}

The fixed point analysis is difficult because one needs to compute 
the solution of the system (\ref{fix}). One can instead examine
qualitatively the evolution of $B$ given by
\be\label{Bt}
B_t = \alpha \rho_{B_x} [ 2 (1-T)(1-B) B_x + B_{xx} + (1-B)^2 T_x], \ee
see the Appendix for the derivation. This shows that as long as 
$\rho_{B_x}$  remains zero $B$ i.e. for $B_x < J_c$, then $B$ does not change. 
When $\rho_{B_x} \neq 0$, three terms contribute to the evolution of $B$: the first term
in the bracket is a convection term, the second is a diffusion term. These
two terms correspond to a Burger's front type dynamics \cite{Whitham}, 
\cite{cs07} for
the magnetic field entering the sample. The third term is unexpected
and gives a global evolution of $B$. This term couples the
$B$ and $T$ evolution equations and is responsible for
the flux jumps.

From a practical point of view, the model we analyzed allows to analyze 
how different
types of pulses $B_e(t)$ affect the total remanent magnetization of a sample.
A recent study by Moroz et al \cite{moroz21} suggests that 
the magnetization of a sample is larger for trapezoidal pulses than
for triangular pulses.
They used Monte-Carlo simulations for the study. Here we confirm 
their findings using our framework. Fig.  \ref{trapez} shows the 
$M(B_e)$ curve for a triangular pulse of 
duration $t_p=1000$ (continuous line, red online) together with trapezoidal
pulses 1 and 2 with the same gradient and plateau regions of duration 
1000 and 2000 respectively. We observe a relaxation in these plateau regions
for $B_e=0.5$ where $B(x,t)$ goes from $B_C(x)$ with $C>0$ to $B_0(x)$.
This relaxation allows for the larger magnetization observed for the
trapezoidal pulses.
\begin{figure}[H]
\centerline{
\epsfig{file=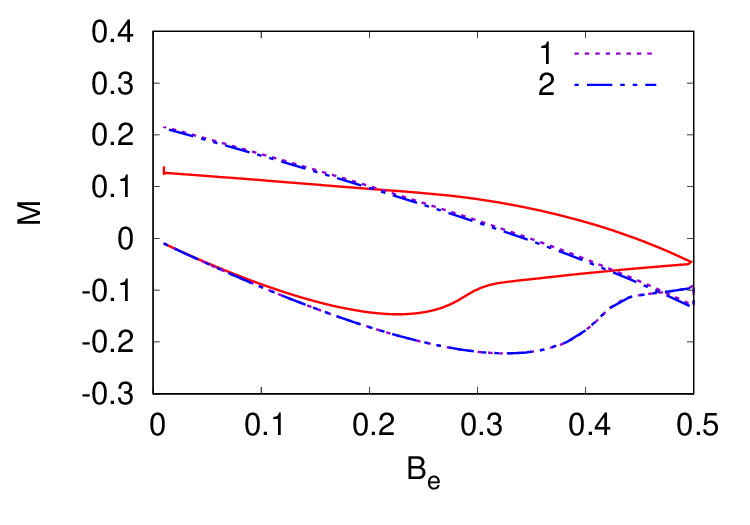, height= 5 cm, width = 12 cm, angle=0}
}
\caption{Magnetization  $M(B_e)$ for a triangular pulse of
duration $t_p=1000$ (continuous line, red online) and trapezoidal pulses
1 and 2 with plateaux duration $1000$ and $2000$ respectively.}
\label{trapez}
\end{figure}

Another direction for increasing the total remanent magnetization is to
send in several pulses and adapt the temperature, see for
example the protocol followed by Fujishiro et al \cite{fuji07}.
The authors send a first pulse, then lower the temperature and send a
second pulse. They observe that the magnetization 
increases from the first pulse to the second. 
Fig. \ref{twopulse} shows $M(B_e)$ for a first triangular
pulse in continuous line (red online) with
$B_{max}=0.5, ~T_e=0.5$ and $t_p=1000$. Three continuations
were tested: (a) $B_{max}=0.6, ~T_e=0.3$, (a) $B_{max}=0.6, ~T_e=0.5$
and (c) $B_{max}=0.7, ~T_e=0.3$. The protocol (a)
gives the largest magnetization in accordance
with the findings of \cite{fuji07}.
It is interesting to observe that 
protocol (c) with $B_{max}=0.7$ does not
increase magnetization.
\begin{figure}[H]
\centerline{
\epsfig{file=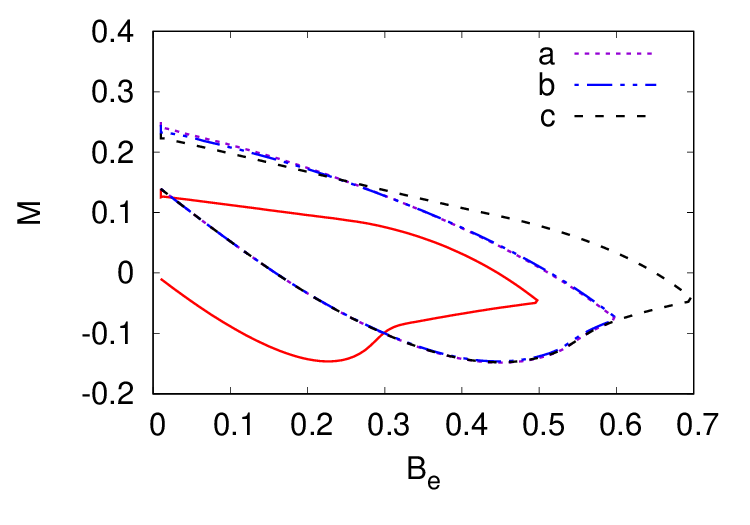, height= 5 cm, width = 12 cm, angle=0}
}
\caption{Magnetization  $M(B_e)$ for a triangular pulse of
duration $t_p=1000$ (continuous line, red online) and three continuations:
(a) $B_{max}=0.6, ~T_e=0.3$, (a) $B_{max}=0.6, ~T_e=0.5$
and (c) $B_{max}=0.7, ~T_e=0.3$.}
\label{twopulse}
\end{figure}

\subsection{Conclusion}

We analyzed mathematically and numerically a flux-jump model 
for type II superconductors in a 1D configuration where 
the magnetic field is a scalar $B$. 
The model involves a nonlinear diffusion equation for $B$ coupled
to a diffusion equation for the temperature $T$. In addition
the critical current $J_c$ 
depends on the magnetic field $B$ and the temperature $T$.

We determined the typical time scales of the Joule heating, magnetic
field dynamics and temperature diffusion for MgB$_2$ and YBaCuO
from the literature. Non dimensionalizing the equations using 
the Joule heating time, we reduced the problem to two parameters
and could understand better the 
mathematical couplings between the different fields. We considered
how the system responds to an incoming magnetic field pulse.

Regarding flux jumps, we describe them using fixed point solutions of
the equations for $B$ and $T$. These solutions $B_0(x)$ and $B_C(x)$
match well the numerical results in the different regimes. 
More globally, the dynamics of $B$ is given by an inhomogeneous Burger's 
convection-diffusion equation whose solutions are fronts. These fronts 
have an irregular speed because of the inhomogeneities. In 
addition, there is a nonlinear driving term $(1-B)^2 T_x$  and this
is responsible for the flux jumps.
We observed that flux jumps occur for pulses of duration
$$t_p \approx t_{mag} < t_{diff},$$
mostly at low temperature and
medium magnetic fields, which is consistent with the form of the
driving term. 
Flux jumps are always preceded by large 
increases in $T$ and $T_x$. Their position depends on the rate
of increase ${d B_e \over dt}$ and they are not affected by 
the duration of the pulse if it is large enough.

To address the practical question of making better pulse charged 
magnets, we varied the incoming pulse duration $t_p$ in different conditions.
For medium temperatures, we found that large $t_p$ maximize the trapped field
and can cause flux jumps. If the pulse amplitude is too large, the
sample turns normal and any trapped field quickly dissipates with
time. At low temperatures, the heat capacity is no longer constant
so that the $T$ equation becomes nonlinear. Here, 
again for large duration pulses, we observe larger
flux jumps and slightly larger trapped fields. Interestingly, sending
another pulse on the sample does not change significantly the magnetization.

A future study could be to understand the nature of this "trapped field
state" and what are the parameters that affect it. 
In a more realistic setting, one could
look for an adapted sample geometry to maximize this trapped field.

\section*{Acknowledgements}
The authors thank J. Durrell for useful discussions.
N. R. acknowledges the support of the Normandie Regional Government and the
European Union through grant "Supramag". The authors thank the 
Centre R\'egional Informatique et d'Applications Num\'eriques de 
Normandie for the use of it's computational ressources.

\appendix 
\section{Fixed point of the equations and oscillations of $B$}

If the heat capacity is large and the pulse is short, 
one can neglect the evolution of the temperature and assume $T(x)=T_e$.
Then equations (\ref{bt1},\ref{tt1}) decouple and we can look
for a fixed point of equation (\ref{bt1}). This is the object of
this section.

Denoting the partial derivatives as subscripts for simplicity
of writing, we have
$$B_t = \alpha \partial_{x}\left[\rho(B,B_x,T)\right],$$
$$ = \alpha ( \rho_B B_x + \rho_{B_x} B_{xx} + \rho_T T_x) .$$
From 
$$\rho(B_x) = 0.5 \left [1+\tanh({B_x-J_c\over w}) \right ](B_x-J_c), $$
and 
$$J_c = (1-T)(1-B)^2 ,$$
we can calculate the different terms $\rho_B , \rho_{B_x},\rho_{J_c}, \rho_T$ 
and obtain
\begin{align*}
\rho_{B_x} = {0.5 \over w} \left [1-\tanh^2({B_x-J_c\over w}) \right ](B_x-J_c), \\
+ 0.5 \left [1+\tanh({B_x-J_c\over w}) \right ], \\
\rho_{J_c}  = - \rho_{B_x}, \\
\rho_B  = \rho_{J_c} {J_c}_B= \rho_{B_x} 2 (1-T)(1-B), \\
\rho_T  = \rho_{J_c} {J_c}_T= \rho_{B_x}  (1-B)^2 .  
\end{align*}
Then the evolution of $B$ is given by
\be\label{Bt2}
B_t = \alpha \rho_{B_x} [2(1-T)(1-B) B_x + B_{xx} +  (1-B)^2 T_x] . \ee

To obtain the fixed point $B_0(x)$, we need to solve the equation
\be\label{fixed}
B_x = J_c= (1-T) (1-B)^2 \ee
together with the boundary conditions (\ref{bc1}).
From (\ref{fixed}) we get 
$${B_x \over (1-B)^2} = (1-T),$$
Assuming, $T=T_e$ this is a separable differential equation
which can be integrated as
$$\int_B^{B_e} {dB \over (1-B)^2}= \int_x^1 (1-T_e) dx,$$
to yield the final result
\be\label{bofx}
B(x) = 1 - {1 \over (1-B_e)^{-1} -(1-T)(1-x)} . \ee

The second type of fixed point $B_C(x)$ satisfies
$B_x = J_c+ C$ leading to the inhomogeneous Ricatti equation
\be\label{fixed2}
B_x = (1-T) (1-B)^2 +C \ee
Introducing the change of field $D = 1/(1-B)$ we get
$$D_x = (1-T) [ 1 + {D^2 C \over 1-T} ]$$
which can be integrated as
$$ D \sqrt{ C \over 1-T} =\tan ( \sqrt{ C (1-T)} x + C_1 ),$$
where $C_1$ is a constant.
Going back to $B$ we get
$$ B = 1 - \sqrt{ C \over 1-T} {1 \over \tan ( \sqrt{ C (1-T)} (x-1) + C_2} ),$$
where $C_2$ is determined by $B(x=1) = B_e$ leading to the final
result:
\be\label{bk2}
B_C(x) = 1 - \sqrt{C \over 1-T} \tan^{-1}\left [ \sqrt{C (1-T)}(x-1)+ {\rm atan}(\sqrt{C \over 1-T} {1 \over 1-B_e} ) \right ]
. \ee

\end{document}